\documentstyle[12pt]{article}
\begin{document}

\newcommand{\nc}{\newcommand}

\nc{\eqn}[1]{Eq.~\ref{eq:#1}}
\nc{\eqns}[2]{Eqs.~\ref{eq:#1} and \ref{eq:#2}}
\nc{\fig}[1]{Fig.~\ref{fig:#1}}
\nc{\figs}[2]{Figs.~\ref{fig:#1} and \ref{fig:#2}}
\input epsf
\nc{\figdir}{} 
\nc{\figmac}[5]{\begin{figure}
\centerline{\parbox[t]{#1in}{\epsfbox{\figdir #2.ps}}}
\caption[#4]{\label{fig:#3} #5}\end{figure}}
%
\def\epsfsize#1#2{\ifdim#1>\hsize\hsize\else#1\fi}

\renewcommand{\slash}[1]{/\kern-7pt #1}
\newcommand{\beq}{\begin{equation}}
\newcommand{\eeq}{\end{equation}}
\newcommand{\beqn}{\begin{eqnarray}}
\newcommand{\eeqn}{\end{eqnarray}}
\newcommand{\beqns}{\begin{eqnarray*}}
\newcommand{\eeqns}{\end{eqnarray*}}
\newcommand{\nn}{\nonumber}

\def\cA{{\cal A}}
\def\cB{{\cal B}}
\def\cC{{\cal C}}
\def\cD{{\cal D}}
\def\cE{{\cal E}}
\def\cF{{\cal F}}
\def\cG{{\cal G}}
\def\cH{{\cal H}}
\def\cI{{\cal I}}
\def\cK{{\cal K}}
\def\cM{{\cal M}}
\def\cN{{\cal N}}
\def\cO{{\cal O}}
\def\cP{{\cal P}}
\def\cR{{\cal R}}
\def\cS{{\cal S}}
\def\cX{{\cal X}}
\def\cZ{{\cal Z}}

\def\qb{\bar{q}}
\def\kb{\bar{k}}
\def\rb{\bar{r}}
\def\Qb{\bar{Q}}
\def\Kb{\bar{K}}
\def\Rb{\bar{R}}

\def\to{\rightarrow}
\newcommand{\lra}{\leftrightarrow}
\newcommand{\la}{\langle}
\newcommand{\ra}{\rangle}
\def\A#1#2{\la#1#2\ra}
\def\B#1#2{[#1#2]}
\def\s#1#2{s_{#1#2}}
\def\h#1#2#3#4{\la#1#2#3#4\ra}
\def\P#1#2{{\cal P}_{#1#2}}
\def\L#1#2{\left\{#2\right\}_{#1}}
\newcommand{\Lmin}{{\rm L}}
\def\Li{{\rm Li_2}}
\def\ms{$\overline{{\rm MS}}$}

\newcommand{\y}{\gamma}
\newcommand{\yf}{\gamma_5}
\newcommand{\yh}{\hat{\gamma}}
\newcommand{\yt}{\tilde{\gamma}}
\newcommand{\yb}{\bar{\gamma}}
\newcommand{\ellb}{\bar{\ell}}
\newcommand{\g}{g}
\newcommand{\gh}{\hat{g}}
\newcommand{\gt}{\tilde{g}}
\newcommand{\qbar}{\bar{q}}

\newcommand{\al}{\alpha}
\newcommand{\be}{\beta}
\newcommand{\del}{\delta}
\newcommand{\eps}{\epsilon}
\newcommand{\ve}{\varepsilon}
\newcommand{\ib}{{\bar\imath}}
\newcommand{\jb}{{\bar\jmath}}

\newcommand{\as}{\alpha_s}
\newcommand{\bz}{\chi_{BZ}}
\newcommand{\nr}{\theta^*_{NR}}
\newcommand{\asj}{\al_{34}}

\newcommand{\kt}{\tilde{k}}
\newcommand{\Pp}{{\cal P}_+}
\newcommand{\Pm}{{\cal P}_-}
\newcommand{\Ppm}{{\cal P}_\pm}
\newcommand{\Pmp}{{\cal P}_\mp}

\newcommand{\Nf}{N_f}
\newcommand{\Nc}{N_c}
\newcommand{\cg}{c_\Gamma}

\newcommand{\ycut}{y_{\rm cut}}


\begin{titlepage}
\begin{flushright}
SLAC--PUB--7490\\
hep-ph/9705218\\
\end{flushright}
\begin{center}
\vspace*{4cm}
{\large\bf
NEXT-TO-LEADING ORDER  CORRECTIONS TO \newline
$e^+ e^- \to $ FOUR JETS}
\vskip 0.4cm
Adrian Signer\footnote{
Research supported by the Swiss National Science Foundation and
by the Department of Energy under grant DE-AC03-76SF00515}  \\
\vskip 0.1cm
{\it Stanford Linear Accelerator Center, Stanford, CA 94309} \\

\vskip 6cm
\end{center}

\begin{abstract}
\noindent
I present a general purpose Monte Carlo program for calculating the
next-to-leading order corrections to arbitrary four-jet quantities in
electron-positron annihilation. In the current version of the program,
some subleading in color terms are neglected. As an example, I calculate
the four-jet rate in the Durham scheme as well as the
Bengtsson-Zerwas angular distribution at $\cO(\as^3)$ and compare the
results to existing data.     
\end{abstract}

\end{titlepage}
\setcounter{footnote}{0}

\newpage


Four-jet events have been copiously produced at the $Z$-pole and at
smaller energies at electron-positron colliders. These events allow the
study of new aspects of QCD. For instance, the three-gluon coupling and
the dependence on the number of light flavors, $\Nf$, enter already at tree
level. Therefore, this seems to be an ideal place for a measurement of the
color factors \cite{AngDistExp, OPAL95, ALEPH97}. Even if there is no
doubt about the correctness of QCD as 
the theory of the strong interactions, these measurements are not purely
academic. In fact, the ongoing debate about the existence of light
gluinos could be settled immediately by measuring $\Nf$ precisely
enough (see e.g. ref. \cite{ALEPH97} and references therein). Indeed, the
addition of a massless gluino amounts to a change of 
$\Nf$ from 5 to 8. Also at LEP2 four-jet events will play an important
role, since they are the main background for the $W$ pair production.
Thus, a next-to-leading order prediction for such quantities is highly
desirable. 

In this talk I present some next-to-leading order results for the four
jet fraction and some angular distributions. These results have been
obtained with a program \cite{DS} which can compute an  arbitrary four-jet
quantity at next-to-leading order. However, as will be discussed below,
some subleading in color contributions have been neglected so far.

Any next-to-leading order calculation involves basically two major
steps. First, the corresponding one-loop amplitudes have to be computed and
second, the phase-space integration has to be performed. This second step
involves the cancellation of the real and virtual singularities. I used
the general version of the subtraction method as proposed in \cite{FKS} to
do the phase space integrals. Thus, no approximation at all has been made
in this part of the calculation.

The one-loop amplitudes which are needed for the calculation of four-jet
events are $ e^+ e^- \to q \qb q' \qb' $ and $ e^+ e^- \to  q \qb g g $.
Recently, the amplitudes for four quarks in the final state have been
computed \cite{GloverMiller, Zqqqq}. However, the situation
concerning the one-loop amplitude for the
$q \qb g g $ final state is less satisfactory. Unfortunately, only the
leading in color terms are known so far \cite{Zqqgg}. As a result,
subleading in color pieces of the cross section can not yet be computed at
next-to-leading order. Writing the color decomposition of any four-jet
cross section as 
\beq
\sigma_{4-{\rm jet}}^{1-{\rm loop}} 
= \Nc^2 (\Nc^2 - 1) \left[
 \sigma_4^{(a)} + \frac{\Nf}{\Nc}\, \sigma_4^{(b)} 
 + \frac{\Nf^2}{\Nc^2} \, \sigma_4^{(c)} 
 + \frac{\Nf}{\Nc^3} \, \sigma_4^{(d)} 
 + \cO \left( \frac{1}{\Nc^2} \right)  \right]\, 
\eeq
I calculate $\sigma_4^{(a,b,c,d)}$.

Besides the one-loop amplitudes, the tree-level amplitudes for
the processes $ e^+ e^- \to q \qb q' \qb' g $ and 
$ e^+ e^- \to  q \qb g g g $ are required for the computation of the real
contributions. They have been obtained by several groups 
\cite{FiveJetsBornBGK, FiveJetsBorn}. In the program the results of ref
\cite{FiveJetsBornBGK} are used. 

In the calculation of these amplitudes all quark and lepton masses have
been set to zero. This is usually a very good approximation, although the
$b$-quark mass effects can yield considerable corrections
\cite{MassEffects}. Unfortunately, the complete inclusion of mass effects
at the order $\cO(\as^3)$ is presently out of reach, the main reason being
the fact that the one-loop amplitudes are known only for the massless
case. 

Besides the mass effects and subleading in color terms, three
more contributions were neglected, although in principle their inclusion
would not pose a problem. 

(1) Contributions coming from Pauli exchange. The corresponding
$\cO(\as^2)$ terms are known to 
be numerically tiny and it is expected that the $\cO(\as^3)$ terms are
numerically not very significant. 
\par\noindent

(2) Contributions proportional to the axial coupling $a_q$ of the $Z^0$ 
to quarks.  Analogous terms have traditionally 
been omitted from $\cO(\as^2)$ programs, as they cancel precisely 
between up- and down-type quarks in the final state (for zero quark
mass), and their contribution to the three-jet rate is at the percent
level~\cite{AxialCoupling}.
\par\noindent

(3) Contributions proportional to $(\sum_q v_q)^2$, where $v_q$ is 
the quark vector coupling.   These ``light-by-glue scattering'' 
terms do not appear at $\cO(\as^2)$ at all, have a partial 
cancellation from the sum over quark flavors, and contribute less 
than 1\% to the $\cO(\as^3)$ term in the total 
cross-section~\cite{NNLOTotalCrossSection}. 

I first present the results for the four-jet rate 
$R_4 \equiv \sigma_{4-{\rm jet}}/\sigma_{{\rm tot}}$ at next-to-leading
order in $\as$ in the Durham scheme \cite{Durham}. 
The $\ycut$ dependence is shown in \fig{r4plot}. The solid (dashed) line
represents the one-loop (tree-level) prediction. The renormalization scale
$\mu$ has been chosen to be the center-of-mass energy $\sqrt{s}$, the
number of flavors $\Nf=5$ and $\al_s = 0.118$~\cite{AverageAlphas}.  The
data points are preliminary SLD  data \cite{SLDdata} and are corrected for
hadronization.

 
\figmac{7.0}{MoDu}{r4plot}{}  
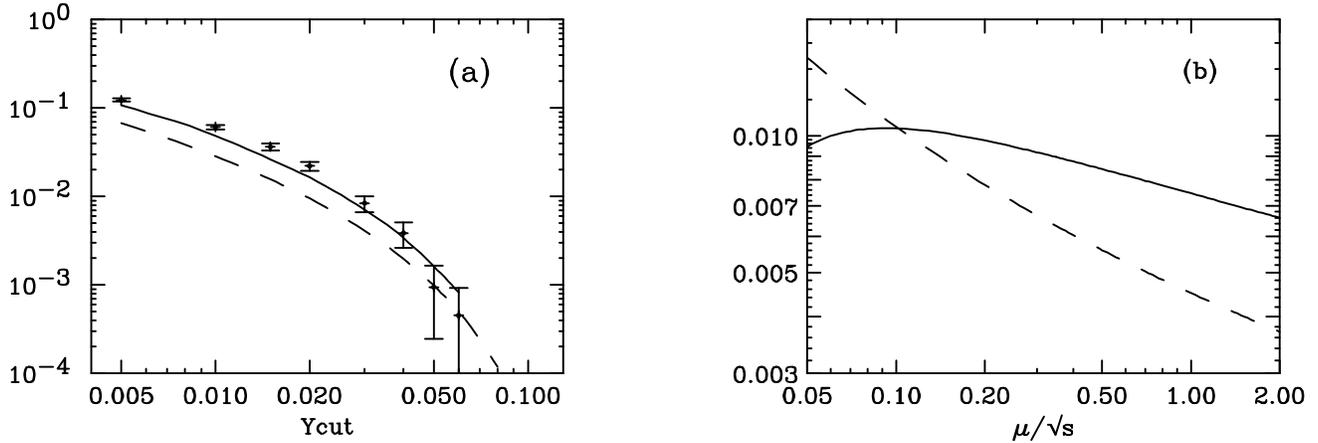
{{\small (a) Solid (dashed) lines represent the one-loop (tree-level) 
predictions for $R_4$ in the Durham scheme for $\mu=\sqrt{s}$ and 
$\al_s = 0.118$. 
(b) Solid (dashed) lines show the dependence of $R_4$
on the renormalization scale $\mu$ 
for the one-loop (tree-level) predictions in the 
Durham scheme, for  $\ycut = 0.03$.
\hfill}}


The truncation of the perturbative expansion for any physical quantity
leads to a dependence of the theoretical prediction on the choice of the
renormalization scale $\mu$. The tree-level $\mu$ dependence is much
stronger for the four-jet rate  than for the three-jet rate, because the
former is proportional to $\as^2$ instead of $\as$. 
As expected, this strong renormalization-scale dependence is reduced
by the inclusion of the next-to-leading order contribution.   
\fig{r4plot} also plots the $\mu$-dependence of $R_4$ at tree-level
and at one-loop for $y_{cut} = 0.03$.

Four-jet angular distributions \cite{FourJetAngles} have been measured by
several collaborations \cite{AngDistExp,OPAL95,ALEPH97}. The general
procedure is to choose a certain jet definition.  Then, in the case of a
four-jet event, the jets are ordered according to their energies such that 
$E_1 > E_2 > E_3 > E_4$. Usually, the most energetic jets  are
associated with the primary quarks whereas the remaining two jets either
origin form a quark or gluon pair (at tree level). This can be exploited
to construct angular variables which have a completely different shape for
the four-quark  and the two-quark-two-gluon final state. Since the two
final states are proportional to different color structures one can attempt
to measure the various color factors and in particular the number of light
flavors $\Nf$. Unfortunately, the four-quark final state is strongly
suppressed. As a result, the full distributions are not very sensitive to
$\Nf$ and the error on the measured value of $\Nf$ is accordingly
large. 

An advantage of the angular distributions lies in the fact that one does
not need to worry about large logarithms coming from a particular choice
of the renormalization scale. The reason for this can easily be understood. 
At tree level, the strong coupling constant appears only in an overall
prefactor.   Since these distributions are normalized, the value of $\as$
and thus the choice of the renormalization scale $\mu$ has no influence at
all on the result. Only the inclusion of the one-loop corrections
introduces a extremely mild $\mu$-dependence.

 
\figmac{4.0}{angles}{ang}{}  
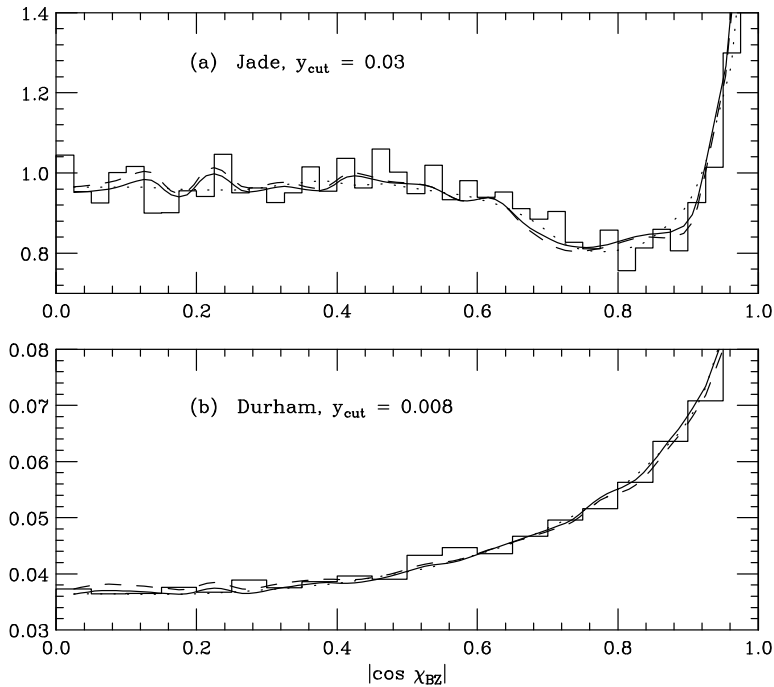
{{ \small Bengtsson-Zerwas distribution at tree level (dotted), one-loop
(solid) and one-loop with $\Nf = 8$ (dashed) compared to (a) OPAL
\cite{OPAL95} and (b) ALEPH \cite{ALEPH97} data (histograms), which are
corrected for detector and hadronization  effects.
\hfill}}


As an example, I consider the Bengtsson-Zerwas angle, $\bz$, and compare
the next-to-leading order prediction with the two most recent analyses of
OPAL and ALEPH \cite{OPAL95,ALEPH97}.  In \cite{OPAL95} jets were defined
according to the JADE scheme with $\ycut = 0.03$, whereas in
\cite{ALEPH97}  the Durham jet algorithm with the E0 recombination scheme
was used and $y_{cut}$ was chosen to be $ 0.008$. Note also that the two
experiments use different normalizations. \fig{ang} compares the 
tree-level (dotted) and next-to-leading order (solid) predictions to the
data which have been corrected for detector and hadronization effects. 
The dotted line can hardly be seen because it nearly coincides with the
solid line. The theoretical curves have been obtained by binning $\bz$
into twenty bins. This rather fine binning results in a somewhat larger
statistical error, which is of the order of 2\% for the shown curves.
In order to illustrate the mild dependence on $\Nf$ I plotted
also the one-loop results for $\Nf = 8$ (dashed). Although this dependence
and thus the precision on the measurement of $\Nf$ may be enhanced by
additional cuts or by $b$-quark tagging \cite{btag}, \fig{ang} shows that
it is very difficult to get a precise measurement of $\Nf$ from angular
distributions alone.


\newpage

\begin{flushleft}
{\bf\large Acknowledgement} \\
\end{flushleft}

It is a pleasure to thank L. Dixon for collaboration in this work and 
P. Burrows, G.~Dissertori and W. Gary for help in comparing the
calculations to data.


\def\np#1#2#3  {{ Nucl. Phys. }{#1}:{#2} (19#3)}
\def\nc#1#2#3  {{ Nuovo. Cim. }{#1}:{#2} (19#3)}
\def\zp#1#2#3  {{ Z. Phys. }{#1}:{#2} (19#3)}
\def\pl#1#2#3  {{ Phys. Lett. }{#1}:{#2} (19#3)}
\def\pr#1#2#3  {{ Phys. Rev. }{#1}:{#2} (19#3)}
\def\prl#1#2#3  {{ Phys. Rev. Lett. }{#1}:{#2} (19#3)}
\def\prep#1#2#3  {{ Phys. Rep. }{#1}:{#2} (19#3)}

\end{document}